\documentclass{scienceadvhack}

\usepackage{scicite}
\usepackage{times}
\usepackage{graphicx}
\usepackage{amsmath}

\newcounter{lastnote}

\title{Accretion-induced variability links young stellar objects, white dwarfs, and black holes}

\author
{Simone~Scaringi,$^{1\ast}$
Thomas~J.~Maccarone,$^{2\dagger}$
Elmar~K{\"o}rding,$^{3\dagger}$
Christian~Knigge,$^{4\dagger}$
Simon~Vaughan,$^{5\dagger}$
Thomas~R.~Marsh,$^{6}$
Ester~Aranzana,$^{3}$
Vikram~Dhillon,$^{7,8}$
Susana~C.~C.~Barros$^{9}$ 
\\
\\
\normalsize{$^{1}$Max-Planck-Institut f{\"u}r Extraterrestriche Physik, D-85748 Garching, Germany}\\
\normalsize{$^{2}$Department of Physics, Texas Tech University, Box 41051, Lubbock, TX 79409–1051, USA.}\\
\normalsize{$^{3}$Department of Astrophysics, Institute for Mathematics, Astrophysics and Particle Physics (IMAPP), Radboud University Nijmegen, P. O. Box 9010, 6500 GL Nijmegen, Netherlands.}\\
\normalsize{$^{4}$School of Physics and Astronomy, University of Southampton, Highfield, Southampton SO17 1BJ, UK.}\\
\normalsize{$^{5}$Department of Physics and Astronomy, University of Leicester, Leicester LE1 7RH, UK.}\\
\normalsize{$^{6}$Department of Physics, University of Warwick, Coventry CV4 7AL, UK}\\
\normalsize{$^{7}$Department of Physics and Astronomy, University of Sheffield, Sheffield S3 7RH, UK}\\
\normalsize{$^{8}$Instituto de Astrofisica de Canarias, E-38205, La Laguna, Tenerife, Spain.}\\
\normalsize{$^{9}$Instituto de Astrof{\'i}sica e Ci{\^e}ncias do Espa{\c c}o, Universidade do Porto, Centro de Astrof{\'i}sica da Universidade do Porto (CAUP), Rua das Estrelas, PT4150-762 Porto, Portugal.}\\
\\
\normalsize{$^\ast$Corresponding author. E-mail: simo@mpe.mpg.de}\\
\normalsize{$^\dagger$These authors contributed equally to this work.}
}

\date{}

\begin{document}

\maketitle 

\begin{abstract}
The central engines of disc-accreting stellar-mass black holes appear to be scaled down versions of the supermassive black holes that power active galactic nuclei. However, if the physics of accretion is universal, it should also be possible to extend this scaling to other types of accreting systems, irrespective of accretor mass, size, or type. We examine new observations, obtained with \textit{Kepler/K2} and ULTRACAM, regarding accreting white dwarfs and young stellar objects. Every object in the sample displays the same linear correlation between the brightness of the source and its amplitude of variability (rms-flux relation) and obeys the same quantitative scaling relation as stellar-mass black holes and active galactic nuclei. We also show that the most important parameter in this scaling relation is the physical size of the accreting object. This establishes the universality of accretion physics from proto-stars still in the star-forming process to the supermassive black holes at the centers of galaxies.
\end{abstract}

\section*{Introduction \\ \\}
Accretion is the process by which most objects in the Universe grow in mass: from young stellar objects (YSOs) still in the star-forming process, through accreting white dwarfs (WDs) and neutron stars (NSs), to stellar-mass black holes (BHs) and supermassive BHs at the center of galaxies [active galactic nuclei (AGN)]. Because the accreting material almost always carries excess angular momentum, the process usually leads to the formation of a rotating disc that slowly transports the material inward. However, we still do not fully understand the physical processes taking place in accretion discs. In particular, it is not clear whether these processes operate in the same way across different types of astrophysical systems. Nevertheless, several phenomenological similarities between accreting systems have been discovered over the past decade (1--12). This suggests that the physics of accretion may indeed be universal, irrespective of accretor mass, size, or type. 

One powerful probe of the accretion process is provided by the brightness variability that is generated in the disc itself. These brightness variations produce observable light curves that are phenomenologically similar across AGN, stellar-mass BHs, and accreting WDs (1, 7, 8). For example, all of these systems -- as well as accreting NSs -- exhibit aperiodic red-noise variability that produces a linear correlation between the root mean square (rms) variability amplitude and the mean flux across a wide range of time scales (9--13). The variability properties are commonly analyzed in frequency space through the power spectral density (PSD). The power density, $P(\nu)$, gives the distribution of power (rms-squared variability amplitude) as a function of frequency, $\nu$ (or, equivalently, time scale $t=1/\nu$). In systems containing high mass flow rate accretion discs -- such as luminous AGN, stellar-mass BHs (in luminous states), and accreting WDs of the nova-like class -- the low frequency (long time scale) PSDs are usually well described by a simple power law, $P(\nu)\propto \nu^{-\alpha}$, with $\alpha \approx 1$. However, at frequencies above a characteristic break frequency, $\nu_b$, the PSDs ``bend'' to a steeper slope ($\alpha\geq2$) (1, 7, 8). The physical process that determines this characteristic frequency is not clear, but the variability probably originates from the inner accretion disc surrounding the accreting object (14--16). In AGN and stellar-mass BHs, this region primarily emits in X-rays; in accretingWDs and YSO, it mainly emits in the optical/ultraviolet (UV).

\begin{figure}[h]
\begin{center}
\hspace*{-1cm}\includegraphics[scale=0.5]{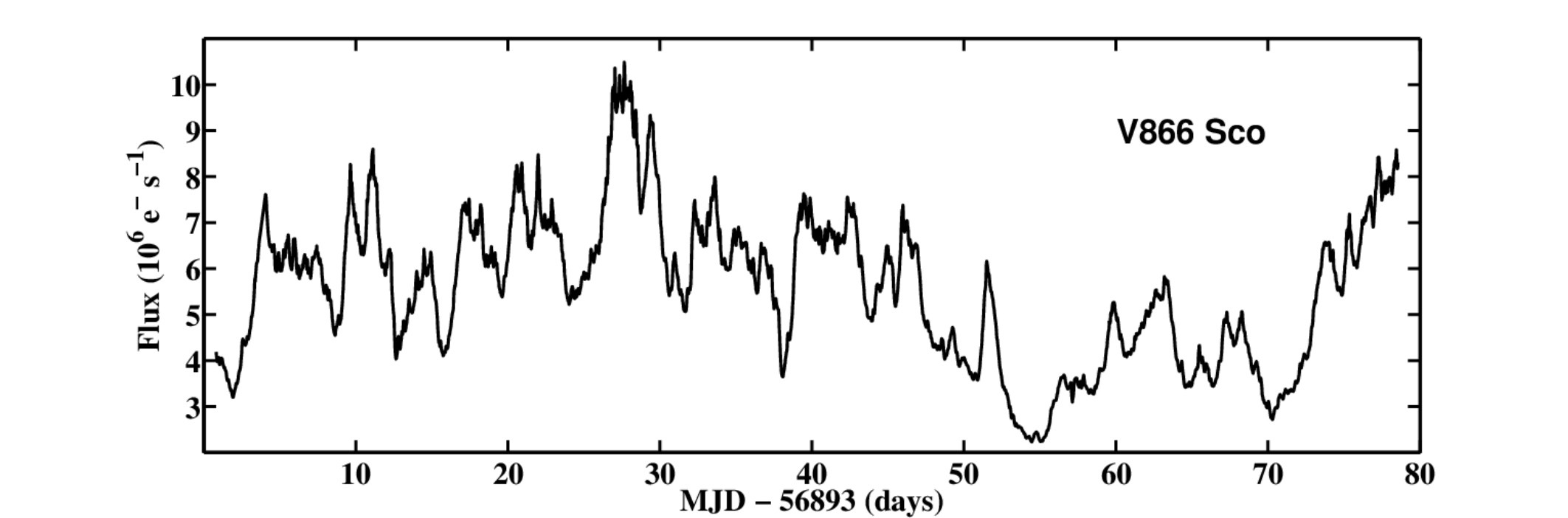}
\caption{
\noindent {\bf \textit{Kepler/K2} light curve of V866 Sco.}
Light curve of the YSO object V866 Sco (EPIC205249328) obtained with the \textit{Kepler/K2} mission during campaign $2$ in long cadence mode (29.4 min). The light curve was extracted from the target pixel files provided by MAST by manually defining large target and background masks.
}
\end{center}
\end{figure}

\begin{figure}[h]
\begin{center}
\hspace*{-0.35cm}\includegraphics[scale=0.46]{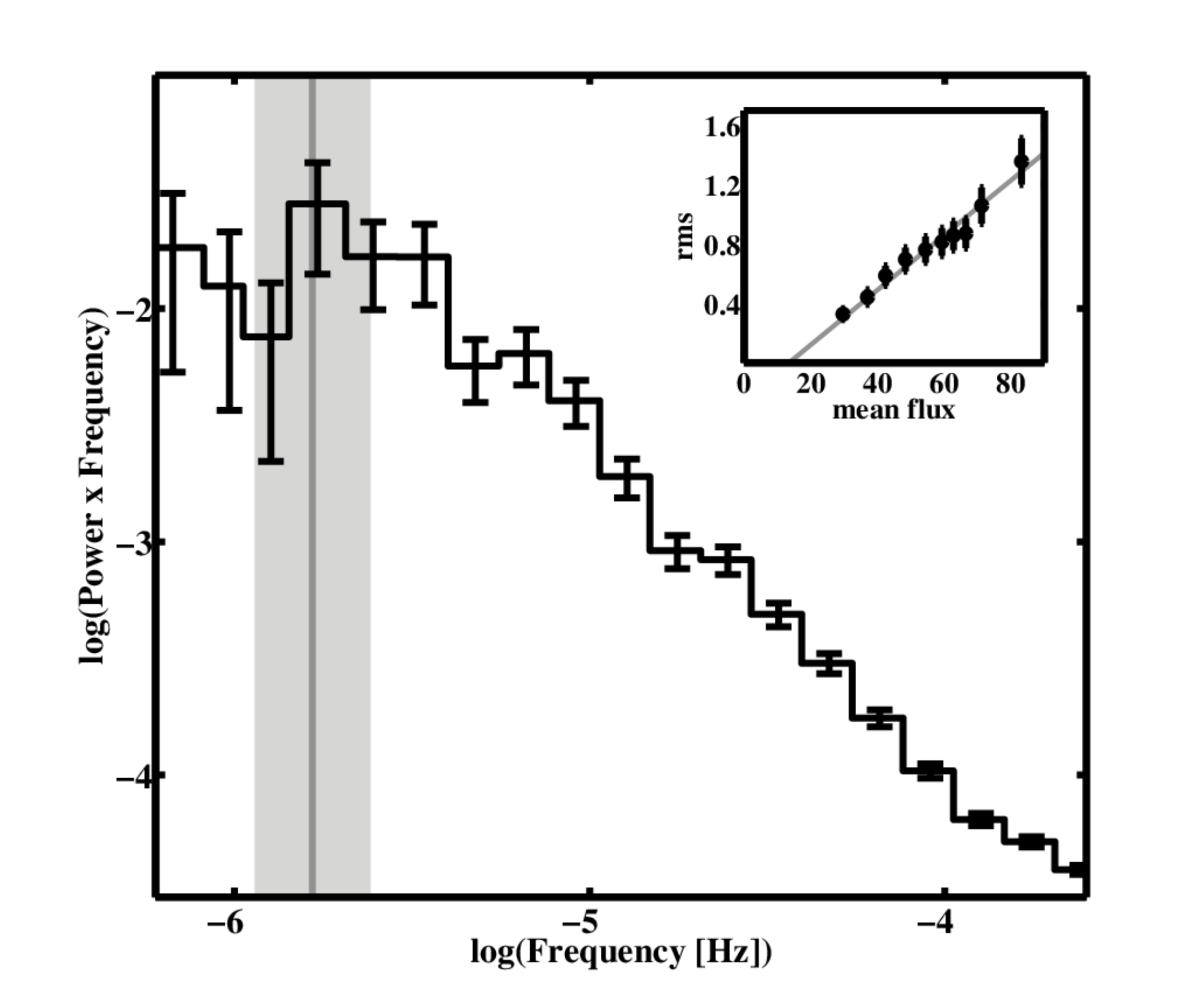}
\caption{
\noindent {\bf Power spectrum and rms-flux relation of the accreting YSO V866 Sco.}
We show the obtained power spectrum of the accreting YSO star V866 Sco (AS 205A or EPIC205249328). The rms-normalized (42) power spectrum was estimated using \textit{Kepler} data from campaign 2, comprising over 78 days of continuous observations with 29.4-min cadence. We mark the measured characteristic bend frequency with the vertical gray solid line, and $1\sigma$ limits with the gray-shaded region (19). The inset shows the linear rms-flux relation from the light curve in units of $10^5$ electrons/s (black data points), computed on 2.5-hour time scales. The gray line is a linear fit to the data. We are able to measure characteristic bend frequencies for the other five YSOs in the \textit{Kepler} sample (which include EPIC204181799, EPIC204395393, EPIC204830786, EPIC204908189, and EPIC205110000), as well as linear rms-flux relations, on a wide range of time scales.}
\end{center}
\end{figure}

\begin{figure}[h]
\begin{center}
\hspace*{-0.5cm}\includegraphics[scale=0.47]{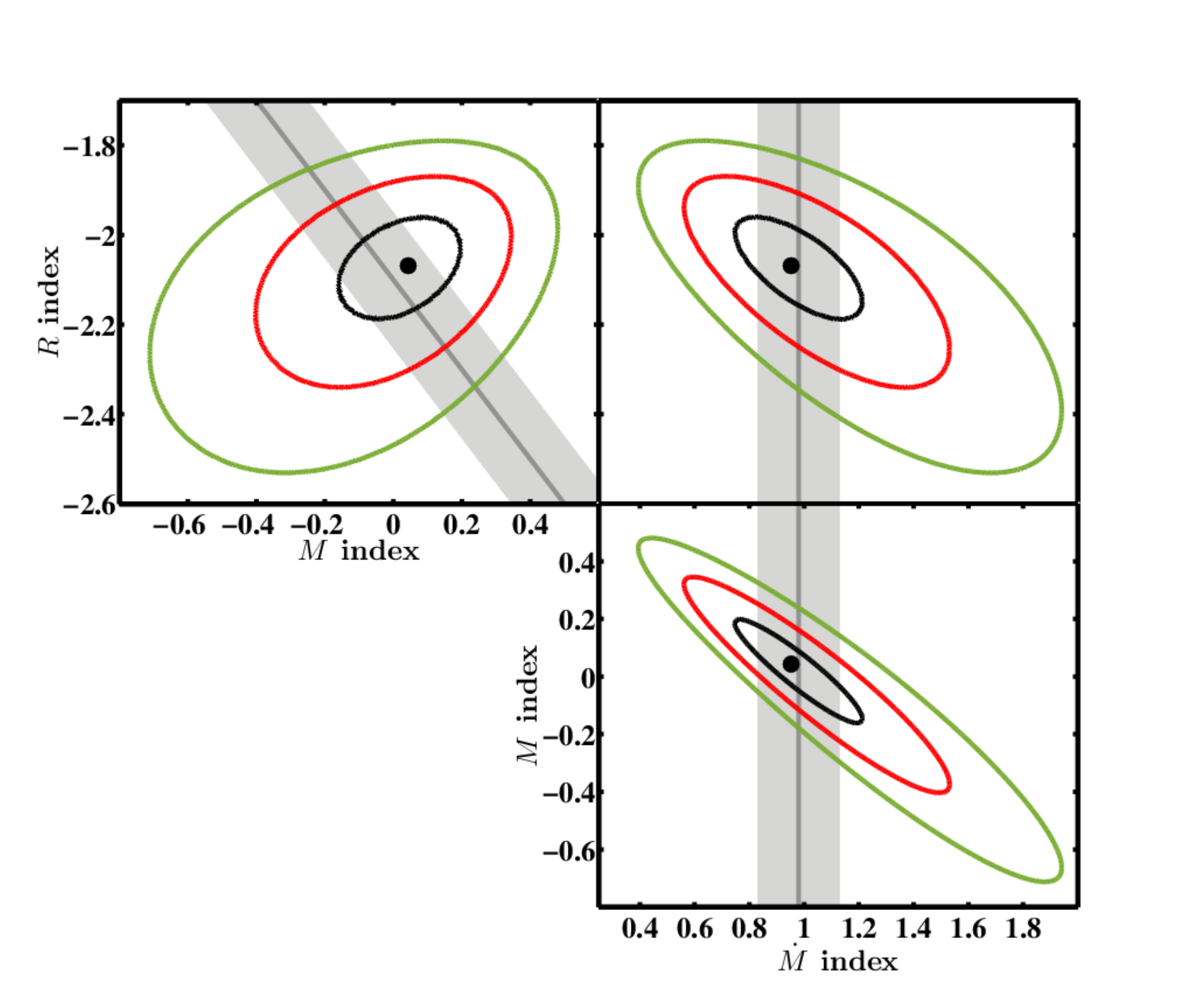}
\caption{
\noindent {\bf Confidence contours for the mass, radius, and mass accretion rate indices.}
We show the confidence contours for the dependency of the characteristic bend frequency on the accretor mass, radius, and mass accretion rate indices. We have fitted the model, $\text{log} \, \nu_b \; = \; A \, \text{log} \, R \; + \; B \, \text{log} \, M \; + \; C \, \text{log} \, \dot{M} \; + D$, where $\nu_b$ represents the measured characteristic bend frequencies, and $R$, $M$, and $\dot{M}$ are the radii, masses, and mass accretion rates, respectively (all using cgs units), to determine the best-fit values for $A$, $B$, $C$, and $D$. The obtained fit is good ($\chi^2 = 38.41$ for $37$ dof) and consistent with the previously obtained fit (1), resulting in $A = −2.07 \pm 0.11$, $B = 0.043 \pm 0.17$, $C = 0.95 \pm 0.22$, and $D = −3.07 \pm 2.61$. The dark gray lines show the previously obtained fit using a BH-only sample (1), with $1\sigma$ errors marked with light gray regions. Given that the previous fit was solely based on BHs, a degeneracy between mass and radius existed, such that it was essentially fitting for $\nu_b \propto R^a M^b$ and the finding that $a + b \approx -2$. The contours refer to the fit including accreting WDs, AGN, and soft-state Galactic BHs and allowing the radius to be an additional free parameter. We thus show that the dominant parameter in the scaling law is a characteristic radius in the inner disc, and not the mass of the accretor as previously thought.}
\end{center}
\end{figure}

\begin{figure}[h]
\begin{center}
\hspace*{-0.5cm}\includegraphics[scale=0.47]{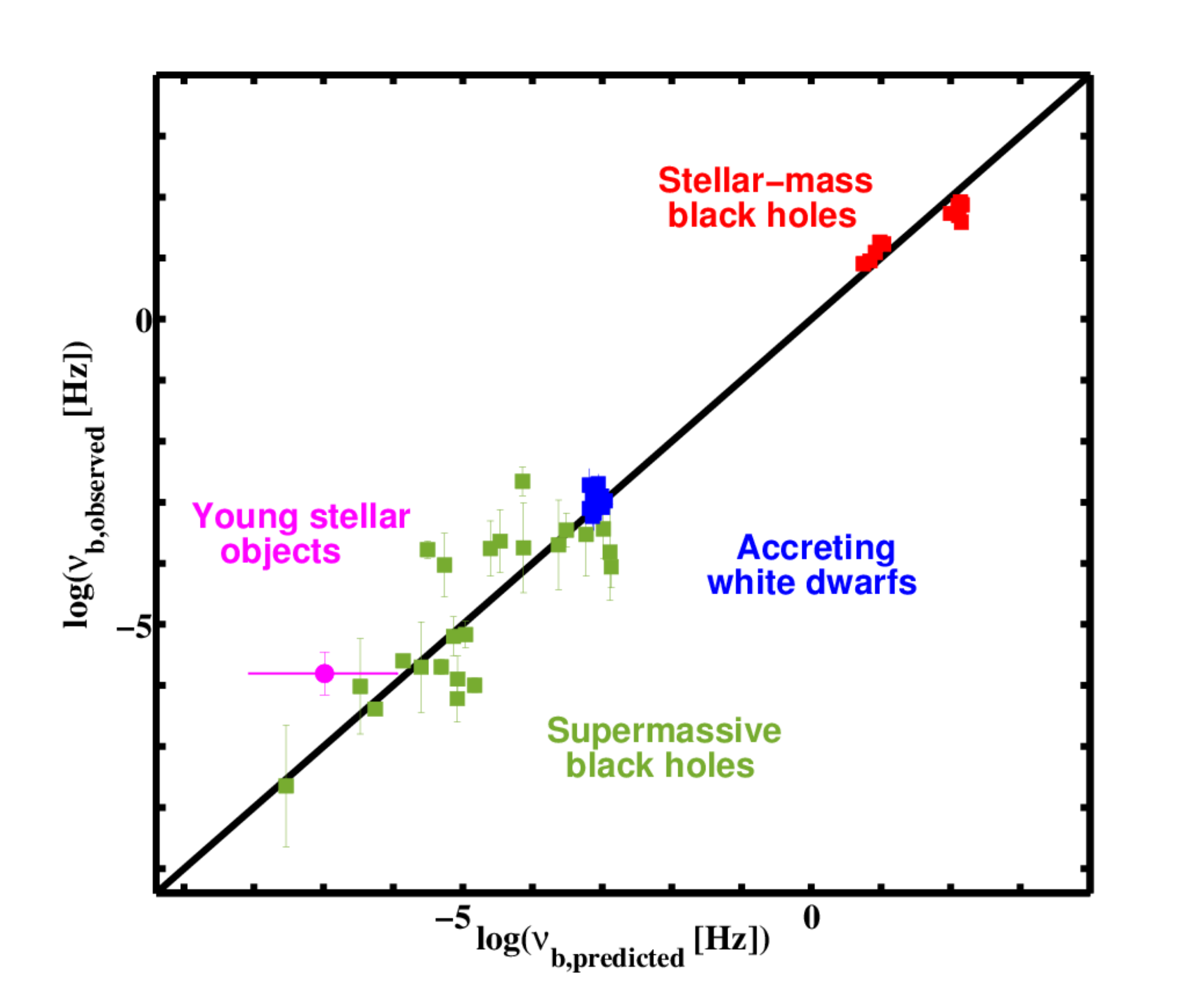}
\caption{
\noindent {\bf Edge-on projection of our sample on the accretion variability plane.}
We show that the predicted characteristic bend frequencies, derived by inserting the observed masses, radii, and mass accretion rates into the best-fit relationship to the combined supermassive BHs, stellar-mass BHs, and accreting WDs, agree very well with the observed characteristic bend frequencies. If the predicted and observed bend frequencies are identical, then an object will exactly lie on the black line. We display supermassive BHs as filled green squares, stellar-mass BHs in their soft state as filled red squares, and accreting WDs as filled blue squares. We additionally show the position of the YSO V866 Sco with the filled magenta circle using the best estimate for its magnetospheric radius at $6.4$ solar radii. The error on the predicted $\nu_b$ assumes strict upper and lower limits for the disc truncation  radius, from $0.1$ to $2$ AU solar radii (30). We thus demonstrate that the variability plane of accreting systems extends from supermassive BHs all the way to YSOs.}
\end{center}
\end{figure}

\section*{Results \\ \\}
Some types of YSOs (pre-main sequence stars accreting from a disc of gas and dust) also show strong signatures of variability in the optical and infrared due to accretion (17). To determine whether the variability properties of noncompact accretors are the same as those of compact ones, we analyzed the \textit{Kepler} data of six YSOs (18) that were observed during campaign 2 of the repurposed \textit{Kepler/K2} mission and that exhibited aperiodic variability. Fig.~1 shows the obtained light curve for one of these objects, the T Tauri star V866 Sco, whereas Fig.~2 shows the resulting PSD and rms-flux relation. The rms-flux relation is clearly linear, and the PSD has a broken power-law shape with an obvious bend at $\nu_b \approx 1.56 \times 10^{−6}$ Hz ($t_b = 1/\nu_b \approx 7.41$ days). The light curves of the other five YSOs in our sample also produce linear rms-flux relations on a wide range of time scales as well as broken power-law PSDs. This strongly suggests that these variability properties are universal among disc-accreting astrophysical systems. 

If this is correct, the characteristic PSD break frequency, $\nu_b$, may be expected to depend quantitatively -- and exclusively -- on the key physical parameters describing such systems. For black holes only, it has already been shown that $\nu_b$ can be predicted quantitatively across eight orders of magnitude in BH mass (1, 3, 4). More specifically, the BH scaling relation can be viewed as a plane in the parameter space given by the (logarithms of) BH mass, accretion rate, and break frequency. However, because the existing relation is derived solely from observations of accreting BHs, it is not clear whether and how it applies to other types of accreting systems. The same limitation has also prevented the unique identification of accretor mass as the underlying key scaling parameter, because, for black holes accreting at high rates, the characteristic inner disc radius and accretor mass linearly scale as $R_{ISCO} \approx 3GM_{BH}/c^2$ (where $G$, $c$, $M_{BH}$, and $R_{ISCO}$ are the gravitational constant, speed of light, BH mass, and the innermost stable circular orbit of a BH with spin parameter $a = 0.8$, respectively; $R_{ISCO} = 6GM_{BH}/c^2$ and $GM_{BH}/c^2$ for nonrotating $a = 0$ and maximally spinning $a = 1$ BHs). 

Motivated by the form of the scaling relation for BHs, which also allows degeneracy between mass and radius in BH samples, we make the ansatz that the break frequency should scale as 

\begin{equation}
\text{log} \, \nu_b \; = \; A \, \text{log} \, R \; + \; B \, \text{log} \, M \; + \; C \, \text{log} \, \dot{M} \; + D
\label{eq:1}
\end{equation}

\noindent if accretion-induced variability is truly universal (where $R$, $M$, and $\dot{M}$ are the characteristic radius, accretor mass, and accretion rate, respectively). To test for the existence of such a universal plane of accretion activity, we expand the existing BH sample by adding a set of eight accreting WDs. All of these belong to the nova-like class, in which the disc extends close to the WD surface. These systems are the key to breaking the BH mass-radius degeneracy because the radius of a WD is three orders of magnitude larger than $R_{ISCO}$ for a comparably massive BH. Accreting NSs also display aperiodic accretion-induced variability, but we do not include them here because their PSDs display multiple breaks and quasi-periodicities (7, 13, 19), which make it difficult to uniquely identify the appropriate break frequency for them. 

For our BH sample, we adopt accretor masses and break frequencies from the literature (1, 20--23) and set the characteristic radius to $R_{ISCO}$ with dimensionless spin parameter $a = 0.8$ (19). BH accretion rates are estimated from measured bolometric luminosities, $L_{bol}$, through the relation $\dot{M} = L_{bol}/\eta c^2$, with $\eta = 0.1$ (24). Our WD sample includes two systems observed with the \textit{Kepler} satellite (MV Lyrae and KIC 8751494), whose variability properties have already been studied previously (8--10).We add to these an additional six accreting WDs (BZ Cam, CM Del, KR Aur, RW Tri, UU Aqr, and V345 Pav) that have been observed with ULTRACAM (19, 25, 26). We use these data sets to estimate break frequencies with errors for all of our accreting WD systems (19). Where possible, we adopt published values for the WD masses (27) and use the theoretical WD mass-radius relationship to obtain the corresponding radii (28). If the WD mass is unavailable, we adopt a value of 0.75 solar masses (29). The mass accretion rate is taken to be $10^{−8}$ solar masses per year for all accreting WDs, which is a typical value for these systems (29). Our results are not sensitive to the exact parameter values we have adopted, and we also allow for conservative errors of 0.4 dex on $\dot{M}$ and 0.2 dex on $M$ and $R$ across the whole sample (19). 

We then perform a grid search in parameter space to determine the best-fit values of $A$, $B$, $C$, and $D$ in Eq.~1 (19). The resulting fit is good [Figs.~3 and 4; see also Materials and Methods (19)] and consistent with a simple generalization of the previously established BH variability plane (1). More specifically, in our new generalized scaling relation, $\nu_b$ is most sensitive to the characteristic radius of the accreting object, rather than the accretor mass. 

To further test the universal nature of the scaling relation, we also show in Fig.~4 the one YSO in our sample -- V866 Sco -- for which published estimates for the mass, accretion rate, and inner disc radius are available (30). The inner discs of some YSOs are thought to be truncated by the magnetosphere of the accreting proto-star. In V866 Sco, the truncation radius has been estimated by fitting the spectral energy distribution (SED) of the system and found to be about $1$ to $10$ times larger than the radius of the proto-star itself (30). We therefore adopt log $R_{\ast} = 0.5 \pm 0.5$ (where $R_{\ast}$ is the stellar radius) as the characteristic radius for this system. Such truncation radius estimates are model-dependent and uncertain, but V866 Sco nevertheless lies close to the universal variability plane. In Fig.~4, the point corresponding to V866 Sco shifts to the right -- and hence closer to the predicted location -- with decreasing inner disc radius. This may suggest that the inner disc radii of YSOs lie at the lower end of the ranges inferred from the SED fits.

\section*{Discussion \\ \\ }
The leading theory for accretion-induced variability in compact objects is the ``fluctuating accretion'' model (14, 15). In this picture, slow accretion rate fluctuations generated in the outer disc propagate inward, where they combine with the progressively faster fluctuations generated at smaller radii. This results in a multiplicative coupling of the variability on all time scales and naturally produces both linear rms-flux relations and broken power-law PSDs (14, 15, 31--33). Our discovery of the same variability characteristics in YSOs establishes the universality of these features across all types of accreting systems, compact or otherwise. This is qualitatively consistent with the fluctuating accretion model because the process it envisages is scale-free (in the sense that it does not rely on a particular system mass or size). 

The universal plane of accretion-induced variability is a further significant step forward in understanding the physics of accretion because it quantitatively links all types of disc-accreting astrophysical systems. For example, the identification of a characteristic inner-disc radius as the dominant parameter -- as opposed to accretor mass or accretion rate -- has obvious implications for the physical interpretation of accretion-induced variability, in general, and the fluctuating accretion model, specifically. Our work thus provides a solid basis for the development of a universal physical model of astrophysical accretion on all scales.

\section*{Materials and Methods}
\subsubsection*{Sample selection}
We attempt to include all relevant sources in our sample. Our sample includes two stellar-mass BHs observed during their soft state (Cyg X$-$1 and GRS 1915$+$105) (3, 7, 22, 34--36) with strong enough variability to measure break frequencies. We also include all unabsorbed, bright AGN with good \textit{XMM-Newton} observations (having more than 40-ks exposures) (21). In addition, we include both nova-like accreting WDs observed during the \textit{Kepler} mission and all six nova-like accreting WDs observed with ULTRACAM (see Table~1 for a full list) in the sample of Barros (25). All systems used are thought to be in an analogous state to the BH (high)soft states, where the accretion discs are thought to extend very close to the accreting objects. Therefore, our assumption that the inner-disc edge of both stellar-mass BHs and AGN extends close to $R_{ISCO}$ is valid. 

Throughout this work, we assume $R_{ISCO}$ for BHs with dimensionless spin parameter $a = 0.8$. Estimates of black hole spins suggest that there is likely to be a range of spin parameter values, but different methods seem to yield different estimates for the spin parameter, at least in some cases (37, 38). Deviations from nonspinning $R_{ISCO}$ estimates will introduce scatter in the scaling relation, but we account for this by introducing systematic uncertainties to our sample. The estimated spins of the two stellar-mass BHs in the sample are both nearly maximal, and the estimated values of the AGN spins that exist for the AGN in the sample are mostly about $a = 0.8$, but most of the AGN in the sample do not have well-established spin estimates (39). If we set $a = 0$ (no spin), we recover slightly different scaling parameters, although they remain consistent with the previously derived mass-scaling relation (1). 

Stellar-mass BHs in their (low)hard state, as well as accreting NSs, have already been shown to seemingly follow the original variability plane (3). However, all hard-state systems seem to be systematically offset from the general soft-state relation. Thus, we only consider soft-state analogs in this work to obtain a refined scaling relation that includes the radius dependence. Soft-state NSs often have very low amplitudes of variability, and much of the variability is coming from the very low frequency noise component, which is often suggested to be due to fusion processes on the NS itself (40). We will address the variability of hard-state systems, including NSs (in all states), in a later work. 

We inspected all YSOs observed during campaign 2 of the \textit{Kepler/K2} mission under the GO program GO2056. These targets have been selected as displaying clear evidence of accretion-induced emission lines (18). The sample contained 71 systems observed in long cadence mode. We extracted light curves for 17 systems from this sample that did not have neighbors close by in the target pixel files provided by the Mikulski Archive for Space Telescopes (MAST). This allowed us to create light curves using relatively large target and background masks, which mitigated the effect of spacecraft jitter in the resulting light curves (41). We visually identified 6 of these 17 light curves as displaying clear aperiodic variability. The light curve obtained for V866 Sco is shown in Fig.~1.

\subsubsection*{PSD fitting}
The PSDs used in this study (both from \textit{Kepler} and ULTRACAM data) were all estimated from evenly sampled sections of data using standard methods (42). Specifically, we computed the rms-normalized periodogram from each continuous section of data, merged the periodograms where appropriate, and averaged the geometrically spaced frequency bins. 

We fit each individual PSD through weighted least squares with a bending power-law of the form

\begin{equation}
P(\nu) \; = \; \frac{R \nu^{\alpha}}{1 + (\frac{\nu}{\nu_b})^{\alpha-\beta}} + N
\label{eq:2}
\end{equation}

\noindent where $P(\nu)$ is variability power as a function of $\nu$; $\alpha$ and $\beta$ are the power-law indices for the low- and high-frequency components, respectively, bending at frequency $\nu_b$; $R$ is a normalization constant; and $N$ is a constant noise component that takes into account high-frequency power not intrinsic to the sources. Confidence intervals were computed by ``bootstrap'' resampling, which was performed 10,000 times to obtain a distribution for the characteristic bend frequencies, $\nu_b$. We set the measured log $\nu_b$ as the mean of the logged distribution and 1$\sigma$ error from its standard deviation.

\subsubsection*{Scaling relation fitting}
The characteristic break frequencies are thought to originate from the inner parts of the accretion disc. These emit most of their radiation in X-rays for AGN and stellar-mass BHs, whereas they emit mostly optical/UV light for accreting WDs and YSOs. Thus, we adopt values of $\nu_b$ for AGN and stellar-mass BHs estimated from X-ray data, whereas we estimate $\nu_b$ for accreting WDs and YSOs from optical data. Although this choice is physically motivated, we note that, for accreting WDs and stellar-mass BHs, optical and X-ray PSDs have already been shown to yield consistent $\nu_b$ for the same object (13, 43). Furthermore, although no such comparison has yet been possible for AGN, the high-frequency power-law slope has been shown to be similar between different AGN at both X-ray and optical wavelengths (44, 45). 

We fit the accretion variability plane with a function of the form taken from Eq. 1 using centimeter-gram-second (cgs) units. We use published values of $\nu_b$ (and errors) and $M$ for AGN and stellar-mass BHs (1, 20--23) and measure our own $\nu_b$ values with errors for accreting WDs and YSOs. In addition, we adopt bolometric luminosities, $L_{bol}$, from the literature and translate these to mass accretion rates $\dot{M} = L_{bol}/\eta c^2$, setting $\eta = 0.1$ (24) for the BH systems. Where possible, we adopt literature values for the masses of accreting WDs (27) and set $M = 0.75$ solar masses otherwise. We also set $\dot{M} =10^{−8}$ solar masses per year for all accreting WDs (29). The uncertainties in $M$, $R$, and $\dot{M}$ are not typically laid out well in the literature, but a general understanding of the scatter in the estimates of these parameters does exist. For the BHs, the scatter in $M$ and $R$ is mostly due to scatter in the $M-\sigma$ relation (46) and uncertainty in the spin distribution, respectively, yielding errors of about 0.2 dex on both parameters. For the WDs, the systems have been taken to have typical values, and the uncertainties represent the breadth of the distribution of these values. Although the spread in $M$ found in accreting WDs is about 0.14 dex (47), we adopt a slightly more conservative value of 0.2 dex accounting for the fact that the population used to obtain the measured spread on $M$ contains very few accreting WDs of the nova-like class, which we use exclusively. The uncertainties in $\dot{M}$ for BHs are mostly driven by uncertainties in the source radiative efficiencies (spin-dependent) and bolometric corrections (48, 49), which we find to be around 0.4 dex. For WDs, the uncertainties in $\dot{M}$ are driven by uncertainties in the distances to the sources (50), which require assuming that the novalike accreting WDs all have similar luminosities, resulting in a 0.4 dex error on $\dot{M}$. We therefore simply adopt a constant 0.2 dex for the uncertainties in $M$ and $R$ and a constant 0.4 dex for the uncertainty in $\dot{M}$ (because this is the hardest to estimate accurately) for the whole sample. We note that modifying the errors on the sample does not affect the obtained fitted parameters for the scaling relation, but rather change the size of the obtained contour levels (Fig.~3). 

To determine the best-fit values for the coefficients $A$, $B$, $C$, and $D$, we performed a total least squares parameter grid search (using errors on all variables) and determined the minimum value of $\chi^2$

\begin{equation}
\chi^2 \; = \; \Sigma \frac{(\text{log} \, \nu_b - E)^2}{\sigma_{\nu_b}^2 \, + \, A^2 \sigma_{R}^2 \, + \, B^2 \sigma_{M}^2 \, + \, C^2 \sigma_{\dot{M}}^2}
\label{eq:3}
\end{equation}

\noindent where $E$ is the log of the predicted frequency, given by the model (Eq.~1). Here, $\sigma_{\nu_b}$ , $\sigma_R$, $\sigma_M$, and $\sigma_{\dot{M}}$ are the errors on log $\nu_b$, log $R$, log $M$, and log $\dot{M}$, respectively. At least to within our adopted errors, the fit is good ($\chi^2 = 38.41$ for $37$ dof), and the coefficients of the log $R$, log $M$, and log $\dot{M}$ terms are consistent with the previously obtained fit for the BH-only sample (1), resulting in $A = −2.07 \pm 0.11$, $B = 0.043 \pm 0.17$, $C = 0.95 \pm 0.22$, and $D = −3.07 \pm 2.61$. By setting the errors on $M$, $R$, and $\dot{M}$ to larger or smaller values, we recover consistent fit parameters as with our original choice of 0.2 dex on $M$ and $R$ and 0.4 dex on $\dot{M}$, albeit with larger or smaller error contours. Thus, our analysis is robust against the precise choice of adopted errors. We also recall that our best fit assumes BH spins of $a = 0.8$ for the whole sample. Deviations from this assumption using either $a = 0$ or $a = 1$ would affect both $R_{ISCO}$ estimates and the accretion efficiency $\eta$. The difference between these two extremes yields about 0.8 dex difference on the BH predicted model frequencies (Fig.~4).

\begin{table*}[htb]
\centering
\caption{ {\bf Samples and parameters used in this work.} The table lists the objects used in this work. We include the adopted masses, radii, mass accretion rates, and break frequencies, $\nu_b$. Where these values have been taken from the literature, we provide the relevant reference. Where these values have been estimated in this work, we refer to Materials and Methods (19). All systems, except for the YSO V866 Sco, have been used for the fit shown in Figs. 3 and 4.}
\label{tab:1}
\begin{spacing}{1.5}
\begin{tabular}{|l|l|l|l|l|l|l|}
\hline
\textbf{Name} & \textbf{Type} & \textbf{log(Mass) [log($M_\odot$)]} & \textbf{log(Radius) [log($R_\odot$)]} & \textbf{log($\dot{M}$) [log($M_\odot$/yr)]} & \textbf{log($\nu_b$) [log(Hz)]} & \textbf{References} \\ \hline
Cyg X-1       & BH            & 1.17                               & -4.03                                & -8.38                                 & 1.09                          & (3, 22, 34, 35)     \\ \hline
GRS 1915+105  & BH            & 1.18                               & -4.02                                & -7.06                                 & 1.87                          & (3, 7, 23, 36)      \\ \hline
Mrk 335       & AGN           & 7.34                               & 2.14                                 & -1.06                                 & -3.76                         & (21)                \\ \hline
H 0707-495    & AGN           & 6.37                               & 1.17                                 & -1.32                                 & -3.82                         & (21)                \\ \hline
ESO 434-G40   & AGN           & 6.3                                & 1.1                                  & -1.45                                 & -4.06                         & (21)                \\ \hline
NGC 3227      & AGN           & 6.88                               & 1.68                                 & -1.89                                 & -3.64                         & (21)                \\ \hline
KUG 1031+398  & AGN           & 6.6                                & 1.4                                  & -0.93                                 & -3.44                         & (21)                \\ \hline
NGC 4051      & AGN           & 6.13                               & 0.93                                 & -2.19                                 & -3.53                         & (21)                \\ \hline
Mrk 766       & AGN           & 6.57                               & 1.37                                 & -1.67                                 & -3.7                          & (21)                \\ \hline
NGC 4395      & AGN           & 5.55                               & 0.35                                 & -4.38                                 & -2.66                         & (21)                \\ \hline
MCG 06.30.015 & AGN           & 6.71                               & 1.51                                 & -1.9                                  & -3.75                         & (21)                \\ \hline
NGC 5506      & AGN           & 7.46                               & 2.26                                 & -1.75                                 & -3.78                         & (21)                \\ \hline
NGC 6860      & AGN           & 7.59                               & 2.39                                 & -1.22                                 & -4.03                         & (21)                \\ \hline
Akn 564       & AGN           & 6.9                                & 1.7                                  & -0.85                                 & -3.46                         & (21)                \\ \hline
NGC 3516      & AGN           & 7.5                                & 2.3                                  & -1.46                                 & -5.7                          & (21)                \\ \hline
NGC 3783      & AGN           & 7.47                               & 2.27                                 & -1.34                                 & -5.2                          & (21)                \\ \hline
NGC 4151      & AGN           & 7.12                               & 1.92                                 & -2.02                                 & -5.9                          & (21)                \\ \hline
Fairall 9     & AGN           & 8.41                               & 3.21                                 & -0.52                                 & -6.39                         & (21)                \\ \hline
NGC 5548      & AGN           & 7.64                               & 2.44                                 & -0.92                                 & -6.22                         & (21)                \\ \hline
NGC 7469      & AGN           & 7.39                               & 2.19                                 & -1.19                                 & -6                            & (21)                \\ \hline
PKS 0558-504  & AGN           & 8.48                               & 3.28                                 & 0.99                                  & -5.17                         & (21)                \\ \hline
IC 4329A      & AGN           & 8.34                               & 3.14                                 & -0.25                                 & -5.6                          & (21)                \\ \hline
PG 0804761    & AGN           & 8.84                               & 3.64                                 & 0.17                                  & -6.02                         & (1)                 \\ \hline
NGC 3516      & AGN           & 7.63                               & 2.43                                 & -1.49                                 & -5.7                          & (1)                 \\ \hline
NGC 4258      & AGN           & 7.59                               & 2.39                                 & -3.61                                 & -7.65                         & (1)                 \\ \hline
KIC 8751494   & WD            & -0.1                               & -2                                   & -8.00                                 & -2.98                         & (10, 19)            \\ \hline
MV Lyr        & WD            & -0.14                              & -1.98                                & -8.00                                 & -2.89                         & (8, 19)             \\ \hline
BZ Cam        & WD            & -0.26                              & -1.92                                & -8.00                                 & -3.23                         & (25, 19)            \\ \hline
CM Del        & WD            & -0.32                              & -1.9                                 & -8.00                                 & -3.1                          & (25,19)             \\ \hline
KR Aur        & WD            & -0.23                              & -1.93                                & -8.00                                 & -3.18                         & (25, 19)            \\ \hline
RW Tri        & WD            & -0.26                              & -1.92                                & -8.00                                 & -2.94                         & (25,19)             \\ \hline
UU Aqr        & WD            & -0.17                              & -1.96                                & -8.00                                 & -2.7                          & (25, 19)            \\ \hline
V345 Pav      & WD            & -0.12                              & -1.99                                & -8.00                                 & -3.09                         & (25,19)             \\ \hline
V866 Sco      & YSO           & 0.13                               & 0.8                                  & -6.14                                 & -5.81                         & (19, 30)                \\ \hline
\end{tabular}
\end{spacing}
\end{table*}

\bibliography{scibib}

\begin{thebibliography}{99}
\bibitem{1}
 I. M. McHardy, E. Koerding, C. Knigge, P. Uttley, R. P. Fender, Active galactic nuclei as scaled-up Galactic black holes. \textit{Nature} \textbf{444}, 730--732 (2006).

\bibitem{2}
 A. Merloni, S. Heinz, T. Di Matteo, A Fundamental Plane of black hole activity. \textit{Mon. Not. R. Astron. Soc.} \textbf{345}, 1057–1076 (2003).

\bibitem{3}
 E. G. K{\"o}rding, S. Migliari, R. Fender, T. Belloni, C. Knigge, I. McHardy, The variability plane of accreting compact objects. \textit{Mon. Not. R. Astron. Soc.} \textbf{380}, 301--310 (2007).

\bibitem{4}
 I. E. Papadakis, The scaling of the X-ray variability with black hole mass in active galactic nuclei. \textit{Mon. Not. R. Astron. Soc.} \textbf{348}, 207--213 (2004).

\bibitem{5}
 E. K{\"o}rding, M. Rupen, C. Knigge, R. Fender, V. Dhawan, M. Templeton, T. Muxlow, A transient radio jet in an erupting dwarf nova. \textit{Science} \textbf{320}, 1318--1320 (2008).

\bibitem{6}
 C. Carrasco-Gonz{\'a}lez, L. F. Rodr{\'i}guez, G. Anglada, J. Mart{\'i}, J. M. Torrelles, M. Osorio, A magnetized jet from a massive protostar. \textit{Science} \textbf{330}, 1209--1212 (2010).

\bibitem{7}
 T. Belloni, D. Psaltis, M. van der Klis, A unified description of the timing features of accreting X-ray binaries. \textit{Astrophys. J.} \textbf{572}, 392--406 (2002).

\bibitem{8}
 S. Scaringi, E. K{\"o}rding, P.Uttley, P. J. Groot, C. Knigge, M. Still, P. Jonker, Broad-band timing properties of the accreting white dwarf MV Lyrae. \textit{Mon. Not. R. Astron. Soc.} \textbf{427}, 3396--3405, (2012).

\bibitem{9}
 S. Scaringi, E. K{\"o}rding, P. Uttley, C. Knigge, P. J. Groot, M. Still, The universal nature of accretion-induced variability: The rms–flux relation in an accreting white dwarf. \textit{Mon. Not. R. Astron. Soc.} \textbf{421}, 2854--2860 (2012).

\bibitem{10}
 M. Van de Sande, S. Scaringi, C. Knigge, The rms–flux relation in accreting white dwarfs: Another nova-like variable and the first dwarf nova. \textit{Mon. Not. R. Astron. Soc.} \textbf{448}, 2430--2437 (2015).

\bibitem{11}
 P. Uttley, I. McHardy, The flux-dependent amplitude of broadband noise variability in X-ray binaries and active galaxies. \textit{Mon. Not. R. Astron. Soc.} \textbf{323}, L26--L30 (2001).

\bibitem{12}
 P. Uttley, I. McHardy, S. Vaughan, Non-linear X-ray variability in X-ray binaries and active galaxies. \textit{Mon. Not. R. Astron. Soc.} \textbf{359}, 345--362 (2005).

\bibitem{13}
 P. Gandhi, The flux-dependent rms variability of X-ray binaries in the optical. \textit{Astrophys. J.} \textbf{697}, L167--L172 (2009).

\bibitem{14}
 Y. E. Lyubarskii, Flicker noise in accretion discs. \textit{Mon. Not. R. Astron. Soc.} \textbf{292}, 679--685 (1997).

\bibitem{15}
 P. Ar{\'e}valo, P. Uttley, Investigating a fluctuating-accretion model for the spectral-timing properties of accreting black hole systems. \textit{Mon. Not. R. Astron. Soc.} \textbf{367}, 801--814 (2006).

\bibitem{16}
 C. Done, M. Gierli{\'n}ski, A. Kubota, Modelling the behaviour of accretion flows in X-ray binaries. Everything you always wanted to know about accretion but were afraid to ask. \textit{Astron. Astrophys. Rev.} \textbf{15}, 1--66 (2007).

\bibitem{17}
 A. A. Miller, L. A. Hillenbrand, K. R. Covey, D. Poznanski, J. M. Silverman, I. K. W. Kleiser, B. Rojas-Ayala, P. S. Muirhead, S. B. Cenko, J. S. Bloom, M. M. Kasliwal, A. V. Filippenko, N. M. Law, E. O. Ofek, R. G. Dekany, G. Rahmer, D. Hale, R. Smith, R. M. Quimby, P. Nugent, J. Jacobsen, J. Zolkower, V. Velur, R. Walters, J. Henning, K. Bui, D. McKenna, S. R. Kulkarni, C. R. Klein, M. Kandrashoff, A. Morton, Evidence for an FU Orionis-like outburst from a classical T Tauri star. \textit{Astrophys. J.} \textbf{730}, 14 (2011).

\bibitem{18}
 K. L. Luhman, E. E. Mamajek, The disk population of the Upper Scorpius association. \textit{Astrophys. J.} \textbf{758}, 13 (2012).

\bibitem{19}
 Materials and Methods section.

\bibitem{20}
 P. Uttley, I. McHardy, X-ray variability of NGC 3227 and 5506 and the nature of active galactic nucleus ``states''. \textit{Mon. Not. R. Astron. Soc.} \textbf{363}, 586--596 (2005).

\bibitem{21}
 O. Gonz{\'a}lez-Mart{\'i}n, S. Vaughan, X-ray variability of 104 active galactic nuclei. XMM-Newton power-spectrum density profiles. \textit{Astron. Astrophys.} \textbf{544}, 57 (2012).

\bibitem{22}
 J. A. Orosz, J. E. McClintock, J. P. Aufdenberg, R. A. Remillard, M. J. Reid, R. Narayan, L. Gou, The mass of the black hole in Cygnus X-1. \textit{Astrophys. J.} \textbf{742}, 10 (2011).

\bibitem{23}
 J. Greiner, J. G. Cuby, M. J. McCaughrean, An unusually massive stellar black hole in the Galaxy. \textit{Nature} \textbf{414}, 522--525 (2001).

\bibitem{24}
 N. I. Shakura, R. A. Sunyaev, Black holes in binary systems. Observational appearance. \textit{Astron. Astrophys.} \textbf{24}, 337--355 (1973).

\bibitem{25}
 S. Barros, thesis, University of Warwick (2008).

\bibitem{26}
 V. S. Dhillon, T. R. Marsh, M. J. Stevenson, D. C. Atkinson, P. Kerry, P. T. Peacocke, A. J. A. Vick, S. M. Beard, D. J. Ives, D. W. Lunney, S. A. McLay, C. J. Tierney, J. Kelly, S. P. Littlefair, R. Nicholson, R. Pashley, E. T. Harlaftis, K. O’Brien, ULTRACAM: An ultra-fast, triple-beam CCD camera for high-speed astrophysics. \textit{Mon. Not. R. Astron. Soc.} \textbf{378}, 825--840 (2007).

\bibitem{27}
 H. Ritter, U. Kolb, Catalogue of cataclysmic binaries, low-mass X-ray binaries and related objects. \textit{Astron. Astrophys.} \textbf{404}, 301--304 (2003).

\bibitem{28}
 M. Nauenberg, Analytic approximations to the mass–radius relation and energy of zerotemperature stars. \textit{Astrophys. J.} \textbf{175}, 417 (1972).

\bibitem{29}
 C. Knigge, I. Baraffe, J. Patterson, The evolution of cataclysmic variables as revealed by their donor stars. \textit{Astrophys. J. Supp.} \textbf{194}, 28 (2011).

\bibitem{30}
 J. A. Eisner, L. A. Hillenbrand, R. J. White, R. L. Akeson, A. I. Sargent, Observations of T Tauri disks at Sub-AU radii: Implications for magnetospheric accretion and planet formation. \textit{Astrophys. J.} \textbf{623}, 952--966 (2005).

\bibitem{31}
 A. Ingram, M. van der Klis, An exact analytic treatment of propagating mass accretion rate fluctuations in X-ray binaries. \textit{Mon. Not. R. Astron. Soc.} \textbf{434}, 1476--1485 (2013).

\bibitem{32}
 S. Scaringi, A physical model for the flickering variability in cataclysmic variables. \textit{Mon. Not. R. Astron. Soc.} \textbf{438}, 1233--1241 (2014).

\bibitem{33}
 P. S. Cowperthwaite, C. S. Reynolds, Nonlinear dynamics of accretion disks with stochastic viscosity. \textit{Astrophys. J.} \textbf{791}, 126 (2014).

\bibitem{34}
 M. Axelsson, L. Borgonovo, S. Larsson, Probing the temporal variability of Cygnus X-1 into the soft state. \textit{Astron. Astrophys.} \textbf{452}, 975--984 (2006).

\bibitem{35}
 J. Wilms, M. A. Nowak, K. Pottschmidt, G. G. Pooley, S. Fritz, Long term variability of Cygnus X-1. IV. Spectral evolution 1999-2004. \textit{Astron. Astrophys.} \textbf{447}, 245--262 (2006).

\bibitem{36}
 S. Trudolyubov, On the two types of steady hard X-ray states of GRS 1915+105. \textit{Astrophys. J.} \textbf{558}, 276 (2001).

\bibitem{37}
 M. C. Miller, J. M. Miller, The masses and spins of neutron stars and stellar-mass black holes. \textit{Phys. Rep.} \textbf{548}, 1--34 (2015).

\bibitem{38}
 M. Kolehmainen, C. Done, M. D{\'i}az Trigo, The soft component and the iron line as signatures of the disc inner radius in Galactic black hole binaries. \textit{Mon. Not. R. Astron. Soc.} \textbf{437}, 316--326 (2014).

\bibitem{39}
 C. S. Reynolds, Measuring black hole spin using X-ray reflection spectroscopy. \textit{Space Sci. Rev.} \textbf{183}, 277--294 (2014).

\bibitem{40}
 P. Reig, S. van Straaten, M. van der Klis, Timing properties and spectral states in Aquila X-1. \textit{Astropys. J.} \textbf{602}, 918--930 (2004).

\bibitem{41}
 S. Scaringi, T. J.Maccarone, R. I. Hynes, E. Körding,G. Ponti, C. Knigge, C. T. Britt, H. van Winckel, Sco X-1 revisited with Kepler, MAXI and HERMES: Outflows, time-lags and echoes unveiled. \textit{Mon. Not. R. Astron. Soc.} \textbf{451}, 3857--3867 (2015).

\bibitem{42}
 M. van der Klis, Fourier techniques in X-ray timing, in Timing Neutron Stars (Kluwer Academic/Plenum Publishers, New York, 1989), p. 27.

\bibitem{43}
 S. Balman, M. Revnivtsev, X-ray variations in the inner accretion flow of dwarf novae. \textit{Astron. Astrophys.} \textbf{546}, 9 (2012).

\bibitem{44}
 R. F. Mushotzky, R. Edelson, W. Baumgartner, P. Gandhi, Kepler observations of rapid optical variability in active galactic nuclei. \textit{Astrophys. J.} \textbf{743}, L12 (2011).

\bibitem{45}
 R. Edelson, S. Vaughan, M. Malkan, B. C. Kelly, K. L. Smith, P. T. Boyd, R. Mushotzky, Discovery of a $\sim$5 day characteristic timescale in the Kepler power spectrum of Zw 229–15. \textit{Astrophys. J.} \textbf{795}, 2 (2014).

\bibitem{46}
 J. Kormendy, L. C. Luis, Coevolution (or not) of supermassive black holes and host galaxies. \textit{Annu. Rev. Astron. Astrophys.} \textbf{51}, 511--653 (2013).

\bibitem{47}
 M. Zorotovic, M. R. Schreiber, B. T. G{\"a}nsicke, Post common envelope binaries from SDSS. XI. The white dwarf mass distributions of CVs and pre-CVs. \textit{Astron. Astrophys.} \textbf{536}, 16 (2011).

\bibitem{48}
 S. L. Shapiro, S. A. Teukolsky, Black holes, white dwarfs, and neutron stars: The physics of compact objects (Wiley-Interscience, New York, 1983), p. 251.

\bibitem{49}
 R. V. Vasudevan, A. C. Fabian, Piecing together the X-ray background: Bolometric corrections for active galactic nuclei. \textit{Mon. Not. R. Astron. Soc.} \textbf{381}, 1235--1251 (2007).

\bibitem{50}
 B. Warner, Absolute magnitudes of cataclysmic variables. \textit{Mon. Not. R. Astron. Soc.} \textbf{227}, 23--73 (1987).



\end{thebibliography}

\bibliographystyle{Science}


\paragraph*{Acknowledgments:} We wish to thank G. Ponti, R. Baptista, J. Green, and M. Middleton, as well as the anonymous referee, for useful discussions relating to this project. \textbf{Funding:} S.S. acknowledges funding fromthe Alexander von Humboldt Foundation. S.V., C.K., V.S.D., and T.R.M. acknowledge funding fromthe Science \& Technology Facilities Council. E.K. and E.A. acknowledge funding from the Dutch research council under grant no. 2013/15390/EW. S.C.C.B. acknowledges support from Fundacao para a ciencia e tecnologia through the Investigador FCT Contract No. IF/01312/2014. Some of the data presented in this paper were obtained from MAST. STScI is operated by the Association of Universities for Research in Astronomy Inc., under NASA contract NAS5-26555. Support for MAST for non-HST (Hubble Space Telescope) data is provided by the NASA Office of Space Science through grant NNX09AF08G and by other grants and contracts. This paper includes data collected by the \textit{Kepler} mission. Funding for the \textit{Kepler} mission is provided by the NASA Science Mission directorate. \textbf{Author contributions:} S.S. initiated the project; collected, reduced, and analyzed the data; interpreted the results; and prepared the manuscript. C.K., E.K., T.J.M., and S.V. equally contributed to project design and interpretation of the results and commented on, discussed, and edited the manuscript. T.R.M. and V.S.D. also commented on and discussed the manuscript. E.A., T.R.M., V.S.D., and S.C.C.B. observed and reduced the ULTRACAM data. \textbf{Competing interests:} The authors declare that they have no competing interests. \textbf{Data and materials availability:} All \textit{Kepler} and \textit{K2} data can be found through MAST. The raw ULTRACAM data are available on request from V.S.D. (vik.dhillon@sheffield.ac.uk) or T.R.M. (t.r.marsh@warwick.ac.uk).


\vspace{0.3cm}
\noindent
Submitted 27 May 2015 \\
Accepted 17 July 2015 \\
Published 9 October 2015 \\
10.1126/sciadv.1500686

\paragraph*{Citation:}
S.~Scaringi, T.~J.~Maccarone, E.~K{\"o}rding, C.~Knigge, S.~Vaughan, T.~R.~Marsh, E.~Aranzana, V.~S.~Dhillon, S.~C.~C.~Barros, Accretion-induced variability links young stellar objects, white dwarfs, and black holes. \textit{Sci. Adv.} \textbf{1}, e1500686 (2015).

\end{document}